\documentstyle[pra,eqsecnum,aps]{revtex}
\def\r{\hat{\rho}}
\def\s{\hat{\vec{\rm S}}}
\def\sa{\hat{\vec{\rm S}}_{1}}
\def\sb{\hat{\vec{\rm S}}_{2}}
\def\t{\hat{\tau}^{k}_{q}(\hat{\vec{\rm S}})}
\def\tp{\hat{\tau}^{k}_{q'}(\hat{\vec{\rm S}})}
\def\tm{\hat{\tau}^{k}_{-q}(\hat{\vec{\rm S}})}
\def\tpqp{\hat{\tau}^{k'}_{q'}(\hat{\vec{\rm S}})^{\dag}}
\def\ta{\hat{\tau}^{k_{1}}_{q_{1}}(\hat{\vec{\rm S}_{1}})}
\def\tb{\hat{\tau}^{k_{2}}_{q_{2}}(\hat{\vec{\rm S}_{2}})}
\begin{document}
\title{ Spin Distributions for Bipartite Quantum Systems}

\author{ A.R.Usha Devi\thanks{email:buniphy@kar.nic.in}}

\address{ Department of Physics,
Bangalore University, Jnanabharathi Campus\\ Bangalore~560~056, India.}

\maketitle

\begin{abstract}
We carryout a comparative study of spin distributions defined over the sphere 
for bipartite quantum spin assemblies. We analyse Einstein-Podolsky-Rosen-Bohm
(EPRB) spin correlations in a spin-$s$ singlet state using these 
distributions. We observe that in the classical limit of $s\rightarrow\infty$, 
EPRB spin distributions turn out to be delta functions, thus reflecting the perfect anticorrelation property of two spin vectors associated with a spin-$s$ singlet state.
\end{abstract}
\section{Introduction}
Distribution function description of quantum mechanics owes its origin to the 
work of Wigner\cite{Wigner}. This description 
{\em offers a framework in which quantum phenomena can be discussed using as 
much classical language as allowed. It appeals naturally to one's intuition 
and 
 can often provide useful physical insight that can not be easily gained 
through other approaches}\cite{Lee}. Irrespective of  a few   shortcomings, like for 
{\em eg.,} appearance of negative probabilities, advantage of using 
the distribution function approach is in the fact that it involves 
classical functions as opposed to operators. Besides the phase space 
distribution functions, where position $q$ and momentum $p$ are the 
statistical variables, there have been several attempts to construct the spin 
distribution functions([3]-[6]).
%\cite{Stratonovich}, \cite{Arecchi}, \cite{Scully},\cite{Ramachandran}. 
Infact, {\em Wigner like} distribution functions have 
been used ([5],\cite{Agarwal}) to discuss EPRB spin correlations in a 
spin-$s$ singlet state. In this  formulation, the spin correlations are cast 
in a structurally similar form to those of local hidden variable models but 
with negative distribution functions. 

In this paper we extend the spin distribution functions defined over the 
sphere {viz.,} the $P-,\ Q-, \ {\rm and}\ F-$ functions  ([4],[6]) 
 for bipartite spin systems characterized by a mixed state 
spin density matrix $\hat{\rho}$ . In Section II we discuss the 
expansion of the spin density matrix in terms of Fano statistical tensor 
parameters\cite{Fano}. In Section III we express the $P-,\ Q-, \ {\rm and}
\ F-$ spin distributions in terms of  Fano statistical tensor parameters. 
In Section IV we extend these distributions to bipartite 
spin systems and study 
the EPRB spin correlations using them. We also show that  distribution 
functions characterizing EPRB  spin-$s$ singlet state approach delta 
function in the classical limit of $s\rightarrow \infty$. 

\section{Fano statistical tensor parameters} 

A set of $(2s+1)^{2}$ spherical tensor operators 
$\{ \t , \ k=0,\ 1,\ 2,\ \ldots ,\  2s {\rm \  and\ }  
q=-k,\ -k+1,\ \ldots , \ k\, \}$ constructed out of the 
spin operator $\s$ through\cite{Rose} 
\begin{eqnarray}\t=&{\cal N}_{sk}\, \left( \s\cdot\vec{\nabla} \right)^{k}
\left\{ r^{k}\,Y_{kq}(\theta,\phi)\right\};\cr
{\cal N}_{sk}=&\displaystyle\frac{2^{k}}{ k!}\left[ \frac{4\pi\, (2s-k)!\, 
(2s+1)}{ (2s+k+1)!}\right]^\frac{1}{ 2},\end{eqnarray}
provides a linearly independent, orthonormal basis and  
any operator, acting on the  $(2s+1)$ dimensional Hilbert 
space of a spin-$s$ assembly, could be resolved into its 
irreducible components in terms of this basis. Here, 
$Y_{kq}(\theta,\phi)$ denote spherical harmonic functions and 
the normalisation constants ${\cal N}_{sk}$ are chosen so 
as to be consistent with the Madison convention\cite{Satchler}: 
\begin{equation}\left<  sm'\right\vert\hat{\tau}^{k}_{q}(\hat{\vec{\rm S}})\left\vert sm\right>
=\sqrt{2k+1}\ c(sks;mqm'), \end{equation}
where $c(sks;mqm')$ denote Clebsch-Gordan coefficients. From 
the property 
\begin{equation}\displaystyle\sum_{m=-s}^{s}\, c(sks;mqm)=(2s+1)\, \delta_{k\,0}\,\delta_{q\,0}\end{equation}
of the Clebsch-Gordan coefficients, it could be easily verified that 
  $\t$ are traceless for all non zero values of $k$ and   
 $\hat{\tau}^{0}_{0}(\s)$ is a $(2s+1)\times (2s+1)$ dimensional 
unit matrix. 

By construction, $\t$ are irreducible under rotations, with 
the transformation  property:
\begin{eqnarray}\t \left. {\rm rotation}\atop 
\longrightarrow \right. \left[\t \right]^{\rm rotated} =&
\hat{R}(\alpha,\beta,\gamma)\, \t \hat{R}^{\dag}
(\alpha,\beta,\gamma)\cr 
=&\displaystyle\sum_{q'=-k}^{k}\, D^{k}_{q'\,q}(\alpha,\beta,\gamma)\,\tp ,
\end{eqnarray} 
where $D^{k}$ denotes $(2k+1)$ dimensional irreducible representation 
of rotations\cite{Rose2} and  $\alpha,\ \beta,\ \gamma$ are the Euler angles 
of rotation. Hermiticity property of $\t$ is 
expressed through, 
\begin{equation}\t^{\dag}=(-1)^{q}\,\tm .\end{equation}
The orthonormality property, 
\begin{equation} {\rm Tr}\left( \t\, \tpqp \right)=(2s+1)\, \delta_{k\,k'}
\delta_{q\,q'},\end{equation}
could be readily realised by making use of Eq.(2.2), (2.5) and the 
properties,
\begin{eqnarray}c(s_{1}s_{2}s;m_{1}m_{2}m)=&(-1)^{s_{1}-m_{1}}
\sqrt{\frac{2s+1}{2s_{2}+1}}\ c(s_{1}ss_{2};m_{1}-m-m_{2}),\cr
c(s_{1}s_{2}s;m_{1}m_{2}m)=&(-1)^{s_{1}+s_{2}-s}
\ c(s_{1}s_{2}s;-m_{1}-m_{2}-m),\cr
c(s_{1}s_{2}s;m_{1}m_{2}m)=&(-1)^{s_{1}+s_{2}-s}
\ c(s_{2}s_{1}s;m_{2}m_{1}m),\end{eqnarray}
together with the orthogonality   
\begin{equation}\displaystyle\sum_{m_{1}=-s_{1}}^{s_{1}}
\displaystyle\sum_{m_{2}=-s_{2}}^{s_{2}}
c(s_{1}s_{2}s;m_{1}m_{2}m)\, c(s_{1}s_{2}s';m_{1}m_{2}m')
= \delta_{s\,s'}\,\delta_{m\,m'}, \end{equation}
of the Clebsch-Gordan coefficients. 

Any arbitrary operator $\hat{A}(\s)$ can be expressed in terms of 
the irreducible tensor operators $\t$ as
\begin{equation}
\hat{A}=\sum_{k=0}^{2s}\sum_{q=-k}^{k}\ \t^{\dag}\ a^{k}_{q}
\end{equation}
where the spherical componets $a^{k}_{q}$  of the operator $\hat{O}$
are given by 
\begin{equation}
a^{k}_{q}={\rm Tr}\left(\hat{O}\ \t\right).
\end{equation}
Specifically, 
the spin density matrix $\r$, which characterises a spin-$s$
assembly, has the resolution:
\begin{equation}\r=\frac{1}{ 2s+1}\displaystyle\sum _{k=0}^{2s}
\displaystyle\sum_{q=-k}^{k}
\t^{\dag}\ t^{k}_{q}, \end{equation}
and   $t^{k}_{q}=\left\langle \t\right\rangle
={\rm Tr}\left( \r\,\t\right)$ are the well-known Fano 
statistical tensor parameters[8].

Depending on the requirements  of the spin density matrix $\r$, 
the Fano statistical tensor parameters $t^{k}_{q}$ 
satisfy the following properties:
\begin{itemize}
\item{Normalization: ${\rm Tr}\left(\r\right)=1 
\Longrightarrow t^{0}_{0}=1,$}

\item{Hermiticity: $\r^{\dag}=\r \Longrightarrow 
t^{k*}_{q}=(-1)^{q}t^{k}_{-q},$}

\item{Property under rotation:}

\begin{center}
\begin{tabular}{ll}
$\left(\r\right)^{\rm rotated}=$&$\frac{1}{ 2s+1}\displaystyle\sum_{k,\,q} 
\left(\t^{\dag}\right)^{\rm rotated} t^{k}_{q}$\cr

 &=$\frac{1}{ 2s+1}\displaystyle\displaystyle\sum_{k,\,q} 
\t^{\dag}\, \left(t^{k}_{q}\right)^{\rm rotated}$\cr 
&$\Longrightarrow \left(t^{k}_{q}\right)^{\rm rotated}=
\displaystyle\sum_{q'=-k}^{k}D^{k}_{q'q}(\alpha,\beta,\gamma)\,t^{k}_{q'}$.
\end{tabular}
\end{center}
\end{itemize}
For an entangled  spin assembly containing  a pair of subsystems 
with spins $s_{1}$ and $s_{2}$, the above discussion can be 
readily generalised( \cite{Fano2},\cite{Mallesh}), and the spin density matrix 
characterising such a composite bipartite system is given by,
\begin{equation}\r_{12} =\frac{1}{ (2s_{1}+1)(2s_{2}+1)}
\displaystyle\sum_{k_{1}=0}^{2s_{1}}\displaystyle\sum_{q_{1}=-k_{1}}^{k_{1}}
\displaystyle\sum_{k_{2}=0}^{2s_{2}}\displaystyle\sum_{q_{2}=-k_{2}}^{k_{2}} 
t^{k_{1}k_{2}*}_{q_{1}q_{2}}
 \left(\ta\otimes\tb \right),
\end{equation} 
where $t^{k_{1}k_{2}}_{q_{1}q_{2}}$ denote  coupled Fano statistical 
tensor parameters, 
$\sa,\ \sb$ denote the spin operators of the subsystems 1 and 2 respectively. 
The reduced  or the subsystem density matrices are obtained 
by taking the partial traces:
\begin{equation}\r_{1}={\rm Tr}_{2}\left(\r_{12}\right) 
=\frac{1}{ (2s_{1}+1)}\displaystyle\sum_{k_{1}=0}^{2s_{1}}
\displaystyle\sum_{q_{1}=-k_{1}}^{k_{1}} 
\ta^{\dag}\,t^{k_{1}0}_{q_{1}0},\end{equation}
and  
\begin{equation}\r_{2}={\rm Tr}_{1}\left(\r_{12}\right) 
=\frac{1}{ (2s_{2}+1)}\displaystyle\sum_{k_{2}=0}^{2s_{2}}\displaystyle\sum
_{q_{2}=-k_{2}}^{k_{2}} 
\tb^{\dag}\,t^{0k_{2}}_{0q_{2}}. \end{equation}
where we have made use of Eq.(2.2) and  (2.3).
The system is said to be entangled iff 
\begin{equation}\r_{12}\neq \r_{1}\otimes  \r_{2}
\Longrightarrow t^{k_{1}k_{2}}_{q_{1}q_{2}}\neq t^{k_{1}0}_{q_{1}0}
\,t^{0k_{2}}_{0q_{2}}.\end{equation}

\section{$P-\ Q-\ {\rm and}\ F $ distributions in terms of 
Fano statistical tensor parameters.}

The coherent state description of electromagnetic fields has proved to be 
successful 
 in providing an insight into the relationship between semiclassical and 
quantum theories of light\cite{Glauber}.The diagonal coherent state 
representation or 
$P$-representation for the density matrix,
\begin{equation}\hat{\rho}=\int \, d\alpha 
\,P(\alpha)\left\vert\alpha \right>\left<\alpha \right\vert , 
\end{equation}
has proven to be very useful in bringing many of the results of quantum 
electrodynamics 
into forms similar to those of classical theory and the expectation values of 
any quantum operators $\hat{A}$ could be realised as classical averages in 
terms of the weight function  $P(\alpha)$ through 
\begin{equation}\left< \hat{A}\right>={\rm Tr}\left[ \hat{\rho}\, 
\hat{A}\right]=
\int \, d\alpha \,P(\alpha)\, A(\alpha), \end{equation}
where $A(\alpha)$ is the classical function corresponding to the  operator 
$\hat{A}.$ Arecchi {\it et.al.,}[4] introduced the analogue of the diagonal 
coherent state representation for the spin density matrix $\hat{\rho}$ via the 
relation
\begin{equation}\hat{\rho}=\int d\Omega\, P(\theta,\phi)\, 
\left\vert\theta\phi\right>\left<
\theta\phi\right\vert , \end{equation} where $d\Omega=\sin\theta\, 
d\theta\, d\phi$ and 
and $\left\vert\theta\phi\right>$ represents the spin coherent state(SCS) or 
Bloch state, defined  as a rotated maximum {\it $'$down$\,'$}  spin state 
$\left\vert  s-s \right>:$
\begin{eqnarray}\left\vert\theta\phi\right>=&e^{(\tau\, \hat{\rm S}_{+}
-\tau^{*}\, \hat{\rm S}_{-})}\left\vert   s-s \right> 
= \hat{R}(\phi-\pi,\theta,\pi-\phi)\left\vert  s-s \right> \cr
=&\displaystyle\sum_{m=-s}^{s}\,\sqrt{\left( {2s \atop s+m} \right) }
\left(\cos\frac{\theta }{ 2}\right)^{s-m}\, 
\left(\sin\frac{\theta }{ 2}\right)^{s+m}\,e^{-i(s+m)\phi}
\left\vert sm\right>. \end{eqnarray}  
Here, $\tau=\frac{1}{ 2} \theta\, e^{-i\phi}$ and $\hat{\rm S}_{\pm}$ are the 
spin ladder operators; $\hat{R}(\phi-\pi,\theta,\pi-\phi)$  denotes rotation 
through 
Euler angles $\phi-\pi,\ \theta$ and $\pi-\phi .$ From the normalisation 
condition 
${\rm Tr}\left[ \hat{\rho}\right]=1$ we have
\begin{equation}\int d\Omega\, P(\theta,\phi)=1 \end{equation}
 i.e., the weight function $P(\theta,\phi)$ in the  diagonal spin 
coherent state representation   is a normalised function. With the help of 
$P(\theta,\phi),$  
the quantum expectation value of any arbitrary spin observable $\hat{A}$  is 
given by the classical average 
\begin{equation}\left< \hat{A} \right>=\int d\Omega\, P(\theta,\phi)\, 
A(\theta,\phi),\end{equation}
where  
\begin{equation}A(\theta,\phi)=\left<\theta\phi\right\vert\hat{A}
\left\vert\theta\phi\right>,
 \end{equation}
 is a classical function corresponding to the quantum mechanical 
operator $\hat{A}.$ 
For example corresponding to  the spin operator $\hat{\vec{\rm S}}=
(\hat{\rm S}_{1},\ 
\hat{\rm S}_{2},\ \hat{\rm S}_{3})$ we obtain 
\begin{equation}\vec{\rm S}(\theta,\phi)=\left<\theta\phi\right\vert
\hat{\vec{\rm S}}\left\vert\theta\phi\right>=s\ \vec{ n}(\pi-\theta,\phi),
\end{equation}
where $\vec{n}(\pi-\theta,\phi)\equiv   (\sin\theta\cos\phi,\  
\sin\theta\sin\phi,\ 
-\cos\theta)$  is a unit vector defined on the Bloch  sphere. Thus in the 
$P$-representation of spin, one can visualise the quantum expectation value of 
spin as a classical statistical average of random  orientations of angular 
directions 
$\vec{n}(\pi-\theta,\phi).$ 
 
Observe that the Fano statistical tensors $t^{k}_{q}$  
 can be  expressed as the classical averages 
\begin{equation}t^{k}_{q}={\rm Tr}\left[\r \t\right]=\int\! d\Omega \, 
P(\theta,\phi)\,\left<\theta\phi\right\vert
\t\left\vert\theta\phi\right\rangle. \end{equation}
The expectation values of the irreducible tensor operators $\t$  in the 
spin coherent states $\left\vert\theta\phi\right\rangle$ 
 can be simplified as  follows:
\begin{eqnarray}\left<\theta\phi\right\vert\t\left\vert\theta\phi\right\rangle
=&\left<s-s\right\vert \hat{R}^{\dag}(\phi-\pi,\theta,\pi-\phi)\,
\t \,\hat{R}(\phi-\pi,\theta,\pi-\phi)\left\vert s-s\right\rangle\cr
=&\displaystyle\sum_{q'=-k}^{k}\left<s-s\right\vert 
\hat{\tau}^{k}_{q'}(\hat{\vec{\rm S}}) \left\vert s-s\right\rangle\ 
D^{k*}_{qq'}(\phi-\pi,\theta,\pi-\phi)\cr
=&\displaystyle\sum_{q'=-k}^{k}c(sks;-s0-s)\,\sqrt{2k+1}\,\delta_{q'\ 0}
\,D^{k*}_{qq'}(\phi-\pi,\theta,\pi-\phi)\cr
=&(-1)^{k}c(sks;s0s)\,\sqrt{4\pi}\ Y_{kq}(\theta,\phi-\pi),
\end{eqnarray}
where we have made use of the transformation property of $\t$ under 
rotations. Utilising the explicit expression for the Clebsch-Gordan 
coefficient $c(sks;s0s)$\cite{Varshalovich} given by, 
\begin{equation}c(sks;s0s)=(2s)!\left[\frac{(2s+1)}{ (2s-k)!(2s+k+1)!}
\right]^{\frac{1}{ 2}},\end{equation}
we obtain,
\begin{equation}\left<\theta\phi\right\vert\t\left\vert\theta\phi
\right\rangle=
\sqrt{4\pi}(-1)^{k+q}(2s)!\left[\frac{(2s+1)}{ (2s-k)!(2s+k+1)!}
\right]^{\frac{1}{ 2}}
\, Y_{kq}(\theta,\phi),\end{equation}
where we have used the symmetry[11] $Y_{kq}(\theta,\phi-\pi)=(-1)^{q}
\,Y_{kq}(\theta,\phi),$ 
of the spherical harmonic functions. Substituting Eq.(3.12) in Eq.(3.9), 
we observe that $P(\theta,\phi)$ must be of the form 
\begin{equation}
P(\theta,\phi)= \frac{1}{\sqrt{ 4\pi}}
\sum_{k=0}^{2s}\sum_{q=-k}^{k}(-1)^{k+q} {\cal P}_{sk}
\, t^{k}_{q}\, Y_{kq}^{*}(\theta,\phi),
\end{equation}
in order to reproduce the Fano statistical tensors $t^{k}_{q}$ in this 
representation.  The coefficients $ {\cal P}_{sk}$  are given by,  
\begin{equation}
 {\cal P}_{sk}=\frac{1}{ (2s)!}\sqrt{\frac{(2s-k)!
(2s+k+1)! }{ (2s+1)}}.
\end{equation}

Another useful distribution function which can be derived using SCS is the 
positive, normalised $Q$-function[2, 5]:
\begin{equation} 
Q(\theta,\phi)=\frac{(2s+1)}{ 4\pi}\left< \theta \phi
\right\vert\hat{\rho}\left\vert\theta\phi\right>. 
\end{equation}
Using Eq.(2.11) and (3.10) we obtain, 
\begin{eqnarray}
Q(\theta,\phi)=& \displaystyle\frac{(2s+1)}{ 4\pi}\, 
\left\langle\theta\phi\right\vert\r\left\vert\theta\phi\right\rangle=
\displaystyle\frac{1}{ 4\pi}\displaystyle\sum_{k=0}^{2s}
\displaystyle\sum_{q=-k}^{k}\ t^{k\,*}_{q}
\,\left<\theta\phi\right\vert\t\left\vert\theta\phi\right\rangle\cr 
=& \displaystyle\frac{1}{ 4\pi}\displaystyle\sum_{k=0}^{2s}
\displaystyle\sum_{q=-k}^{k}\sqrt{4\pi }\ t^{k\,*}_{q}
\,(-1)^{k+q}\,Y_{kq}(\theta,\phi)\, (2s)!\,
\sqrt{\frac{(2s+1)} {(2s-k)!(2s+k+1)!}}\cr
=& \displaystyle\frac{1}{\sqrt{ 4\pi}}
\displaystyle\sum_{k=0}^{2s}\displaystyle\sum_{q=-k}^{k}
\,(-1)^{k+q}\  {\cal Q}_{sk}\ t^{k}_{q}\,Y_{kq}^{*}(\theta,\phi)\, 
\end{eqnarray}
where 
\begin{equation}
{\cal Q}_{sk}=(2s)!\sqrt{\frac{(2s+1)}{(2s-k)!(2s+k+1)!}}.
\end{equation}
It could be readily verified, using the orthonormality 
property of the spherical harmonics[11],
\begin{equation}
\int\! d\Omega \ Y_{kq}(\theta,\phi)\ Y_{k'q'}(\theta,\phi)=
\delta_{k,\ k'}\ \delta_{q,\ q'},  
\end{equation}
that the $P-$ and $Q-$ functions\cite{multipoles} given 
by Eq.(3.13) and Eq.(3.16) respectively, are normalised:
\begin{eqnarray}
\int\! d\Omega \, P(\theta,\phi)=&
\displaystyle\frac{1}{\sqrt{4\pi}} \displaystyle\sum_{k=0}^{2s}
\displaystyle\sum_{q=-k}^{k}
\,(-1)^{k+q}\  {\cal P}_{sk}\ t^{k}_{q}  
\int\! d\Omega \, Y_{kq}^{*}(\theta,\phi)\cr 
=&\displaystyle\frac{1}{\sqrt{ 4\pi}} \displaystyle\sum_{k=0}^{2s}
\displaystyle\sum_{q=-k}^{k}
\,(-1)^{k+q}\  {\cal P}_{sk}\ t^{k}_{q} \ \sqrt{4\pi}\ 
\delta_{k,\ 0}\ \delta_{q,\ 0}=1
\end{eqnarray}
and 
\begin{eqnarray}\int\! d\Omega \, Q(\theta,\phi)=&   
\displaystyle\frac{1}{\sqrt{4\pi}}\displaystyle\sum_{k=0}^{2s}
\displaystyle\sum_{q=-k}^{k}
\,(-1)^{k+q}\  {\cal Q}_{sk}\ t^{k}_{q}  
\int\! d\Omega \, Y_{kq}^{*}(\theta,\phi)\cr 
=&\displaystyle\frac{1}{\sqrt{ 4\pi}} \displaystyle\sum_{k=0}^{2s}
\displaystyle\sum_{q=-k}^{k}
\,(-1)^{k+q}\  {\cal Q}_{sk}\ t^{k}_{q} \ \sqrt{4\pi}\ 
\delta_{k,\ 0}\ \delta_{q,\ 0}=1
\end{eqnarray}
Also, the  reality of the sums 
$\displaystyle\sum_{q=-k}^{k}(-1)^{q}\ t^{k}_{q}\ Y_{kq}^{*}(\theta,\phi)$ 
ensures that $P(\theta,\phi)$ and $Q(\theta,\phi)$ are real.

One can also construct  a  spin distribution function  over the sphere  
 using the characteristic function approach[6]. 
In this method, a classical distribution function $F(\vec{X})$ with 
$\vec{X}\equiv(X_{1},\ X_{2},\ 
X_{3})$ as the associated random variables, is constructed by taking the 
Fourier inverse of its characteristic function $\phi(\vec{I})$, which 
is the expectation value of $e^{i\ \vec{I}\cdot\vec{X}}$ i.e., 
\begin{equation}
\phi(\vec{I})=E\left( e^{i\vec{I}\cdot\vec{X}} \right)=\int\!\int\!\int 
\, d^{3}X\, F(\vec{X})e^{i\vec{I}\cdot\vec{X}}.
\end{equation}
Here, $E\left(\ldots\right)$ denotes the expectation value. 
It can be readily seen that the distribution function $F(\vec{X})$ is 
obtained as  Fourier inverse of the characteristic function $\phi(\vec{I})$: 
\begin{equation}
F(\vec{X})=\frac{1}{ (2\pi)^{3}}\int\!\int\!\int\, d^{3}I\, e^{-i\vec{I}
\cdot\vec{X}}\phi(\vec{I}).
\end{equation}
Making use of the well-known expansion
\begin{equation}
e^{-i\vec{I}\cdot\vec{X}}=4\pi\displaystyle\sum_{k=0}^{\infty}
\displaystyle\sum_{q=-k}^{k}\ 
i^{k}\ j_{k}(IX)\ Y_{kq}^{*}(\theta,\phi )\ Y_{kq}(\theta_{I},\phi_{I}),
\end{equation}
where, $j_{k}$  denotes the spherical Bessel function and  the
  spherical polar co-ordinates of  
$\vec{I}$ and $\vec{X}$ are denoted,  respectively, by 
$(I,\ \theta_{I},\  \phi_{I}),\ \ (X,\  \theta,\  \phi).$ 
One can express the  characteristic function 
$\phi(\vec{I})$ in terms of the spherical moments 
\begin{equation}
\mu_{q}^{k}=E\left[ j_{k}(IX)\,Y_{kq}^{*}(\theta,\phi)\right],
\end{equation}
as, 
\begin{equation}
\phi(\vec{I})=4\pi\displaystyle\sum_{k=0}^{\infty}
\displaystyle\sum_{q=-k}^{k}\ (i)^{k}\,
Y_{kq}^{*}(\theta_{I},\phi_{I})\, \mu_{q}^{k}.
\end{equation}
In the case of quantum spin assemblies, $\vec{X}$ corresponds to spin operators
 $\s\equiv (\hat{S}_{1},\ \hat{S}_{2},\ \hat{S}_{3})$ and 
the classical average $E\left(\ldots\right)$ is replaced by 
the quantum mechanical average ${\rm Tr}\left(\r\ldots\right)$. 
Moreover, a spin distribution is expected to automatically reflect 
the constancy 
of the squared angular momentum i.e.,  $\hat{S}^{2}=\hat{S}_{1}^{2}+
\hat{S}_{2}^{2}+\hat{S}_{3}^{2}=s(s+1)$. On imposing the condition 
$X_{1}^{2}+X_{2}^{2}+X_{3}^{2}=R^{2}=s(s+1)$ and introducing solid harmonics 
${\cal Y}_{kq}(\vec{X})=R^{k}Y_{kq}(\theta,\ \phi)$   in Eq.(3.24) we obtain, 
 
\begin{equation}
\mu^{k}_{q}=j_{k}(IR)\,R^{-k}\,E\left({\cal Y}_{kq}(\vec{X})\right),
\end{equation}
where $j_{k}(IR)\,R^{-k}$ has  been taken outside the expectation value since 
$R=\sqrt{s(s+1)}$ is a constant. We can now use the correspondence rule
\cite{Rose3} 
\begin{equation}{\cal  Y}_{kq}(\vec{X})\longrightarrow 
\displaystyle\frac{1}{k!\
 {\cal N}_{sk}}\, 
\hat{\tau}^{k}_{q}
(\hat{\vec{\rm S}}),
\end{equation} 
so that
\begin{equation}E\left( {\cal Y}_{kq}(\vec{X})\right)=
\displaystyle\frac{1}{k!\ {\cal N}_{sk}}\,  {\rm Tr}
\left[ \hat{\rho}\  \hat{\tau}^{k}_{q}
(\hat{\vec{\rm S}})\right]=
\displaystyle\frac{t^{k}_{q}}{k!\ {\cal N}_{sk}}, \end{equation} 
for $k\leq 2s$  and $E\left( {\cal Y}_{kq}(\vec{X})\right)=0$ for $k\geq 2s,$ 
in accordance with the Wigner-Eckart theorem[11]. Thus the characteristic 
function of Eq.(3.25) takes the form,
\begin{equation}\phi (\vec{I})=4\pi\displaystyle\sum_{k=0}^{2s}
\displaystyle\sum_{q=-k}^{k}\ (i)^{k}\,
\displaystyle\frac{1}{k!\ {\cal N}_{sk}}\,R^{-k}\, j_{k}(IR) 
\, Y_{kq}^{*}(\theta_{I},\phi_{I})\, t_{q}^{k}.
\end{equation} 
The Fourier transform of $\phi (\vec{I})$ can now  be readily obtained by 
making use of the 
result[15],
 \begin{equation}\int_{0}^{\infty}\,\int_{0}^{\pi}\,\int_{0}^{2\pi}\,  
I^{2}\,dI\ \sin\theta_{I}\,d\theta_{I}
 \, d\phi_{I}\ e^{-i\vec{I}.\vec{X}}j_{k}(IR)\,
Y_{kq}^{*}( \theta_{I},\phi_{I})
=\displaystyle\frac{2\pi^{2} }{ R^{2}}(-i)^{k}\,\delta(R-X)\, 
Y_{kq}^{*}( \theta,\phi)
\end{equation}
and  we obtain the spin distribution function over the surface of the sphere 
of radius $R=\sqrt{s(s+1)}:$
 \begin{eqnarray}
F(\vec{X})=&\displaystyle\frac{1}{(2\pi)^{3}} 
\displaystyle\int_{0}^{\infty}\,
\int_{0}^{\pi}\,\int_{0}^{2\pi}\,  I^{2}\,dI\ \sin\theta_{I}\,d\theta_{I}
 \, d\phi_{I}\ e^{-i\vec{I}.\vec{X}}\, \phi (\vec{I})\cr
& =\delta(R-X)\,\displaystyle\sum_{k=0}^{2s}
\displaystyle\sum_{q=-k}^{k}\, 
\displaystyle\frac{1}{k!\ {\cal N}_{sk}}\, R^{-k-2}
\, t^{k}_{q}\, Y_{kq}^{*}( \theta,\phi).
\end{eqnarray}
The normalised distribution $F (\theta,\phi)$ of angular co-ordinates 
$\theta,\  \phi$ is obtained through the relation,
\begin{equation}
F(\theta,\phi)=\displaystyle\int_{0}^{\infty}X^{2}\,dX\, 
F (\vec{X})
=\displaystyle\sum_{k=0}^{2s}\displaystyle\sum_{q=-k}^{k} {\cal F}_{sk}\ 
t^{k}_{q}\ Y_{kq}^{*}(\theta,\phi )
\end{equation}
where 
\begin{equation}
 {\cal F}_{sk}=\frac{1}{ 2^{k}}
\left[ \frac{(2s+k+1)!}{ 4\pi (2s-k)!(2s+1)\{s(s+1)\}^{k}}
\right]^{\frac{1}{ 2}}.
\end{equation}
It could be easily verified that $F (\theta,\phi)$ is a real, 
normalised function of $\theta,\ \phi$. All the three  functions 
$P(\theta,\phi), \ Q(\theta,\phi),$ and $F (\theta,\phi)$  have the 
properties of statistical distribution functions in the sense that, they are 
normalised to unity and they yield correct expectation values for quantum 
mechanical spin observables, based on the correspondence rules  associating  
$\t$ to spherical harmonic functions $Y_{kq}(\theta,\phi),$ each  with 
different weight factors, given explicitly through, 
\begin{itemize}
\item{$P$-representation:}
\begin{equation}\t\longrightarrow \sqrt{\frac{4\pi }{ (2s+1)}}\ (2s+1)!
\frac{(-1)^{k+q}\ Y_{kq}(\theta,\phi)}{\sqrt{(2s-k)!(2s+k+1)!}},
\end{equation}
\item{$Q$-representation:}
\begin{equation}\t\longrightarrow \sqrt{\frac{4\pi }{ (2s+1)}}\ 
\frac{(-1)^{k+q}}{ (2s)!}\sqrt{(2s-k)!(2s+k+1)!}\ Y_{kq}(\theta,\phi), \end{equation}
\item{$F $-representation:}
\begin{equation}\t\longrightarrow 2^{k}\left[\frac{4\pi\, (2s+1) 
(2s-k)!\{s(s+1)\}^{k}}{ (2s+k+1)!}\right]^{\frac{1}{ 2}}
Y_{kq}(\theta,\phi).\end{equation}
\end{itemize}

\section{$P-,\ Q-, {\rm\ and}\ F -$ functions for 
bipartite quantum systems}

The $P$ and $Q$ distribution functions   can be  readily 
extendend for bipartite spin assemblies  as follows:
\begin{equation}
\r=\int\!\int\! d\Omega_{1}\  d\Omega_{2}\ 
P(\theta_{1},\phi_{1};\theta_{2},\phi_{2})
\left\vert \theta_{1},\phi_{1};\theta_{2},\phi_{2}\right>
\left< \theta_{1},\phi_{1};\theta_{2},\phi_{2}\right\vert 
\end{equation}
and 
\begin{equation}
Q(\theta_{1},\phi_{1};\theta_{2},\phi_{2})=
\frac{(2s_{1}+1)(2s_{2}+1)}{ (4\pi)^{2}} 
\left< \theta_{1},\phi_{1};\theta_{2},\phi_{2}\right\vert \r 
\left\vert \theta_{1},\phi_{1};\theta_{2},\phi_{2}\right> 
\end{equation}
where $\left\vert \theta_{1},\phi_{1};\theta_{2},\phi_{2}\right>$ are product 
spin coherent states with $(\theta_{1},\phi_{1})$,  $ (\theta_{2},\phi_{2})$ 
representing angular  directions on respective Bloch  spheres corresponding 
to systems with spin $s_{1}$ and $s_{2}$. The results of section  III
 could be easily generalised to the  case  of a pair of systems and  we obtain 
\begin{eqnarray}
P(\theta_{1},\phi_{1};\theta_{2},\phi_{2})=& \displaystyle\frac{1 }{ 4\pi}
\displaystyle\sum_{k_{1}=0}^{2s_{1}}\displaystyle\sum_{q_{1}=-k_{1}}^{k_{1}}
\displaystyle\sum_{k_{2}=0}^{2s_{2}}\displaystyle\sum_{q_{2}=-k_{2}}^{k_{2}}
(-1)^{k_{1}+q_{1}}\ 
(-1)^{k_{2}+q_{2}}\ 
 {\cal P}_{k_{1}s_{1}}\  {\cal P}_{k_{2}s_{2}}\cr
& \times t^{k_{1}k_{2}}_{q_{1}q_{2}}\  Y_{k_{1}q_{1}}^{*}(\theta_{1},\phi_{1})
\ Y_{k_{2}q_{2}}^{*}(\theta_{2},\phi_{2})
\end{eqnarray}
and 
\begin{eqnarray}
Q(\theta_{1},\phi_{1};\theta_{2},\phi_{2})= &\frac{1 }{ 4\pi}
\displaystyle\sum_{k_{1}=0}^{2s_{1}}\displaystyle\sum_{q_{1}=-k_{1}}^{k_{1}}
\displaystyle\sum_{k_{2}=0}^{2s_{2}}\displaystyle\sum_{q_{2}=-k_{2}}^{k_{2}}
(-1)^{k_{1}+q_{1}}\ 
(-1)^{k_{2}+q_{2}}\ 
 {\cal Q}_{k_{1}s_{1}}\  {\cal Q}_{k_{2}s_{2}}\cr 
& \times t^{k_{1}k_{2}}_{q_{1}q_{2}}\  Y_{k_{1}q_{1}}^{*}(\theta_{1},\phi_{1})
\ Y_{k_{2}q_{2}}^{*}(\theta_{2},\phi_{2})
\end{eqnarray}

Let us now consider a EPRB spin-$s$ singlet. The spin density operator in  
this case  is $\r=\left\vert (ss)00\right>\left<(ss)00\right\vert$,  the 
matrix elements of which are  given by 
\begin{equation}
\left\langle sm_{1}';sm_{2}'\vert\r\vert sm_{1};sm_{2}\right\rangle
=\frac{(-1)^{m_{1}-m_{1}'}}{(2s+1)} 
\, \delta_{m_{1}'\,-m_{2}'}\ \delta_{m_{1}\,-m_{2}},
\end{equation} 
in the  $(2s+1)\times (2s+1)$ dimensional direct product spin space of 
particles 1 and 2.
The  coupled Fano statistical parameters $t^{k_{1}k_{2}}_{q_{1}q_{2}}$ 
characterising  EPRB spin correlations are given by, 
\begin{eqnarray}
t^{k_{1}k_{2}}_{q_{1}q_{2}}=&{\rm Tr}[\r\ \ta\times\tb]\cr
=& \displaystyle\sum_{m_{1},m_{1}',m_{2},m_{2}'=-s}^{s}
\frac{(-1)^{m_{1}-m_{1}'}}{(2s+1)} 
\, \delta_{m_{1}'\,-m_{2}'}\ \delta_{m_{1}\,-m_{2}}\ 
\left\langle sm_{1}\vert \ta\vert sm_{1}'\right\rangle
\left\langle sm_{2}\vert \tb\vert sm_{2}'\right\rangle\cr 
 =&\displaystyle\sum_{m_{1},m_{1}'=-s}^{s}
\frac{(-1)^{m_{1}-m_{1}'}}{(2s+1)}
\left\langle sm_{1}\vert \ta\vert sm_{1}'\right\rangle
\left\langle s-m_{1}\vert \tb\vert s-m_{1}'\right\rangle\cr 
 =& \displaystyle\sum_{m_{1},m_{1}'=-s}^{s}
\frac{(-1)^{m_{1}-m_{1}'}}{(2s+1)}\ c(sks;m_{1}'q_{1}m_{1})\ 
c(sks;-m_{1}'q_{1}-m_{1})\sqrt{(2k_{1}+1)(2k_{2}+1)}\cr
&=(-1)^{k_{1}+q_{1}}\,\delta_{k_{1}\,k_{2}}\ 
\delta_{q_{1}\,-q_{2}}
\end{eqnarray}
where we have made use of Eqs. (2.2), (2.7) and (2.8)). 
Thus, the $P$ and $Q$  functions assume the simple form 
\begin{equation}
P(\theta_{1},\phi_{1};\theta_{2},\phi_{2})= \frac{1}{ 4\pi}
\displaystyle\sum_{k=0}^{2s}(-1)^{k} \ {\cal P}_{sk}^{2}\ 
\displaystyle\sum_{q=-k}^{k} Y_{k\,q}^{*}(\theta_{1},\phi_{1})
Y_{k\,q}(\theta_{2},\phi_{2})\end{equation}
and 
\begin{equation}
Q(\theta_{1},\phi_{1};\theta_{2},\phi_{2})=
\frac{1 }{ 4\pi}
\displaystyle\sum_{k=0}^{2s}(-1)^{k}\ Q^{2}_{sk}\ 
\displaystyle\sum_{q=-k}^{k} Y_{k\,q}^{*}(\theta_{1},\phi_{1})
Y_{k\,q}(\theta_{2},\phi_{2})
\end{equation}
for EPRB spin correlations.

The $F$  representation can be generalised for pairs of vector statistical 
variates  $\vec{X}_{1}$ , $\vec{X}_{2}$  denoting $'${\it classical spin 
vectors}$'$ constrained by the conditions $\vert \vec{X}_{1}\vert
=\sqrt{s_{1}(s_{1}+1)}$, $\vert \vec{X}_{2}\vert=\sqrt{s_{2}(s_{2}+1)}$ 
and we obtain,  on using the method outlined in Section III,   
\begin{eqnarray}F(\vec{X}_{1},\vec{X}_{2})=&\delta(\sqrt{s_{1}(s_{1}+1)}-\vert \vec{X}_{1}\vert)
\displaystyle\sum_{k_{1}=0}^{2s_{1}}
\displaystyle\frac{1}{k_{1}!\ {\cal N}_{k_{1}s_{1}}}\, 
\{s_{1}(s_{1}+1)\}^{\frac{-k_{1}-2 }{ 2}} \cr
&  \times\delta(\sqrt{s_{2}(s_{2}+1)}-\vert \vec{X}_{2}\vert)
\displaystyle\sum_{k_{2}=0}^{2s_{2}}
\displaystyle\frac{1}{k_{2}!\ {\cal N}_{k_{2}s_{2}}}\, 
\{s_{2}(s_{2}+1)\}^{\frac{-k_{2}-2 }{ 2}} \cr 
& \hskip 1in \times \displaystyle\sum_{q_{1}=-k_{1}}^{k_{1}}\displaystyle\sum_{q_{1}=-k_{1}}^{k_{1}}
t^{k_{1}k_{2}}_{q_{1}q_{2}}\  Y_{k_{1}q_{1}}^{*}(\theta_{1},\phi_{1})
Y_{k_{2}q_{2}}^{*}(\theta_{2},\phi_{2})\end{eqnarray}
and the normalised distribution of angular variables   i.e., 
$F(\theta_{1},\phi_{1};\theta_{2},\phi_{2})$ is obtained through, 
\begin{eqnarray}
F(\theta_{1},\phi_{1};\theta_{2},\phi_{2})=&
\displaystyle\int_{0}^{\infty}\!\int_{0}^{\infty}\! X_{1}^{2}\, 
dX_{1}\, X_{2}^{2}\, dX_{2}\,F(\vec{X}_{1},\vec{X}_{2})\cr
=& \displaystyle\sum_{k_{1}=0}^{2s_{1}}{\cal F}_{s_{1}k_{1}}
 \displaystyle\sum_{k_{2}=0}^{2s_{2}}{\cal F}_{s_{2}k_{2}}
\displaystyle\sum_{q_{1}=-k_{1}}^{k_{1}}
\displaystyle\sum_{q_{1}=-k_{1}}^{k_{1}}
t^{k_{1}k_{2}}_{q_{1}q_{2}}\  Y_{k_{1}q_{1}}^{*}(\theta_{1},\phi_{1})
Y_{k_{2}q_{2}}^{*}(\theta_{2},\phi_{2}).
\end{eqnarray}
For EPRB  spin correlations $F(\theta_{1},\phi_{1};\theta_{2},\phi_{2})$ 
reduces to 
\begin{equation}
F(\theta_{1},\phi_{1};\theta_{2},\phi_{2})=\frac{1}{4\pi}
\displaystyle\sum_{k=0}^{2s}(-1)^{k}\ {\cal F}^{2}_{sk}\ 
\displaystyle\sum_{q=-k}^{k} Y_{k\,q}^{*}(\theta_{1},\phi_{1})
Y_{k\,q}(\theta_{2},\phi_{2}).
\end{equation} 
Normalisation property of these functions follows naturally from the 
orthogonality of the spherical harmonics. 
It could be verified  that the marginal distributions for system 1 in each
 of these representations is given by 
\begin{eqnarray}
P(\theta_{1},\phi_{1})=&\int\! d\Omega_{1}P(\theta_{1},\phi_{1};\theta_{2},\phi_{2})=\displaystyle\frac{1}{4\pi},\cr 
Q(\theta_{1},\phi_{1})=&\int\! d\Omega_{1}Q(\theta_{1},\phi_{1};\theta_{2},\phi_{2})=\displaystyle\frac{1}{4\pi},\cr 
F(\theta_{1},\phi_{1})=&\int\! d\Omega_{1}F(\theta_{1},\phi_{1};\theta_{2},\phi_{2})=\displaystyle\frac{1}{4\pi},
\end{eqnarray}
where $d\Omega_{1}=\sin\theta_{1}\ d\theta_{1}\ d\phi_{1}.$ Similarly, for 
system 2, the marginal distributions are realised to be 
$P(\theta_{2},\phi_{2})=\displaystyle\frac{1}{4\pi};\ 
Q(\theta_{2},\phi_{2})=\displaystyle\frac{1}{4\pi};\ {\rm and}\ 
F(\theta_{2},\phi_{2})=\displaystyle\frac{1}{4\pi}$. 
These marginal distributions are 
spherically symmetric and hence correspond to totally random 
orientations $(\theta_{1},\phi_{1})$ and  $(\theta_{2},\phi_{2})$ of the spin 
vectors associated with the system. 

The spin correlations $\left<(\sa\cdot \vec{a})(\sb\cdot \vec{b})\right>$ 
could be evaluated using the above distribution functions as follows:
We make  use of the relations between irreducible tensors of rank  1 and the 
spin components $\hat{S}_{1},\  \hat{S}_{2},\ \hat{S}_{2},$ given by 
\begin{eqnarray}\hat{S}_{1}=&\sqrt{\displaystyle
\frac{s(s+1)}{ 6}}\ \left(\hat{\tau}^{1}_{-1}(\s)-
\hat{\tau}^{1}_{1}(\s)\right)\cr
\hat{S}_{2}=&i\sqrt{\displaystyle
\frac{s(s+1)}{ 6}}\ \left(\hat{\tau}^{1}_{-1}(\s)+
\hat{\tau}^{1}_{1}(\s)\right)\cr
\hat{S}_{3}=&\sqrt{\displaystyle
\frac{s(s+1)}{ 3}}\ \hat{\tau}^{1}_{0}(\s)\end{eqnarray}
which lead to the  correspondence rules (see  Eqs.(3.34-3.36) of Section III) 
for the spin operator $\s$ in the three representations discussed above, as 
\begin{eqnarray}
P-{\rm representation:}&  \   
 \s\longrightarrow \ s\ \vec{n}(\pi-\theta,\phi),\cr
Q-{\rm representation:} 
&\ \s\longrightarrow \ (s+1)\ \vec{n}(\pi-\theta,\phi),\cr
F-{\rm representation:}
&\ \s\longrightarrow \ \sqrt{s(s+1)}\ \vec{n}(\theta,\phi)\end{eqnarray}
where $\vec{n}$ denotes a unit vector specified by the  polar co-ordinates 
in the arguments. Thus, we have  
\begin{eqnarray}\left<(\sa\cdot \vec{a})(\sb\cdot \vec{b})\right>=&s^{2}
\int\!\int\! d\Omega_{1}\ d\Omega_{2}\  P(\theta_{1},\phi_{1};\theta_{2},\phi_{2})
\left(\vec{n}_{1}\cdot \vec{a}\right)\left(\vec{n}_{2}\cdot \vec{b}\right)\cr
=&(s+1)^{2}\int\!\int\! d\Omega_{1}d\Omega_{2}\ Q(\theta_{1},\phi_{1};\theta_{2},\phi_{2})
\left(\vec{n}_{1}\cdot \vec{a}\right)\left(\vec{n}_{2}\cdot \vec{b}\right)\cr
=& \{s(s+1)\}^{2}\int\!\int d\Omega_{1}d\Omega_{2}\ F(\theta_{1},\phi_{1};\theta_{2},\phi_{2})
\left(\vec{n}_{1}\cdot \vec{a}\right)\left(\vec{n}_{2}\cdot 
\vec{b}\right)\end{eqnarray}
which,   on expressing $\vec{n}\cdot\vec{a}=\frac{4\pi}{ 3}\displaystyle\sum_{q=-1}^{1}Y_{1q}(\vec{n})
Y_{1q}^{*}(\vec{a})$ and on using the orthonormality property of spherical 
harmonics, yeilds the well-known result for EPRB spin-$s$  correlations,
\begin{equation}
\left<(\sa\cdot \vec{a})(\sb\cdot \vec{b})\right>=-\frac{s(s+1)}{ 3}\,\cos\theta_{ab}.
\end{equation}
Here, $\cos\theta_{ab}=\vec{a}\cdot\vec{b}$.

One can make use of the 
addition theorem for spherical harmonics[11] 
\begin{equation}
\sum_{q}Y_{kq}(\theta_{1},\phi_{1})Y_{kq}^{*}(\theta_{2},\phi_{2})
=\frac{2k+1}{4\pi}\ P_{k}(\cos\theta_{12}),
\end{equation} 
where $P_{k}$ denote Legendre polynomial of order $k$; 
$\cos\theta_{12}=\cos\theta_{1}\cos\theta_{2}+\sin\theta_{1}\sin\theta_{2}
\cos(\phi_{1}-\phi_{2})$,  so that the spin distributions reduce to 
\begin{eqnarray}
P(\theta_{12})=&\displaystyle\frac{1}{(4\pi)^{2}}
\displaystyle\sum_{k=0}^{2j}(-1)^{k}\ (2k+1)\ {\cal P}_{sk}^{2}
\ P_{k}(\cos\theta_{12})\cr
Q(\theta_{12})=&\displaystyle\frac{1}{(4\pi)^{2}}
\displaystyle\sum_{k=0}^{2j}(-1)^{k}
\ (2k+1)\ {\cal Q}_{sk}^{2}\ P_{k}(\cos\theta_{12})\cr
F(\theta_{12})=&\displaystyle\frac{1}{(4\pi)^{2}}
\displaystyle\sum_{k=0}^{2j}(-1)^{k}
\ (2k+1)\ {\cal F}_{sk}^{2}\ P_{k}(\cos\theta_{12}).
\end{eqnarray}
We give below the $P$, $Q$, and $F$ functions explicitly for the 
spin-$\frac{1}{ 2}$ case:
\begin{eqnarray}P^{\frac{1}{ 2}}(\theta_{12})=&
\displaystyle\frac{1}{ (4\pi)^{2}}( 1-9\,\cos\theta_{12})\cr
Q^{\frac{1}{ 2}}(\theta_{12})=&
\displaystyle\frac{1}{ (4\pi)^{2}}( 1-\cos\theta_{12})\cr
f^{\frac{1}{2}}(\theta_{12})=&
\displaystyle\frac{1}{ (4\pi)^{2}}( 1-3\,\cos\theta_{12})\end{eqnarray}
where $\cos\theta_{12}=\cos\theta_{1}\,\cos\theta_{2}+\sin\theta_{1}\,\sin\theta_{2}\,
\cos (\phi_{1}-\phi_{2})$. In Fig.1 we have plotted $P$, $Q$, and $F$ 
functions for spin values $s=\frac{1}{ 2},\ 1,\ \frac{3}{ 2},\ 2.$ 
It could be observed that the distributions show prominent peaks around 
$\theta_{12}=180^{\circ}$ indicating the anticorrelation property of 
associated classical vectors $\vec{n}_{1}$ and $\vec{n}_{2}$. 

Expressing $(2s\pm k)!=(2s)^{2s\pm k}\ 
\displaystyle\prod_{n=\mp k}^{2s-1}\left(1-\frac{n}{2s}\right)$ etc., in 
${\cal P}_{sk},\ {\cal Q}_{sk} {\rm and}\  {\cal F}_{sk}$
 it can be readily observed that 
\begin{equation}
\lim_{s\rightarrow\infty}{\cal P}_{sk}=1,\ \  
\lim_{s\rightarrow\infty}{\cal Q}_{sk}=1,\ \ 
\lim_{s\rightarrow\infty}{\cal F}_{sk}=1,
\end{equation}
which, together with  the completeness property[11] 
\begin{equation}
\sum_{k=0}^{\infty}\sum_{q=-k}^{k}Y_{kq}^{*}(\theta,\phi)\ Y_{kq}(\theta ', 
\phi ')=\delta(\cos\theta -\cos\theta ')\ \delta(\phi-\phi '),
\end{equation} 
and the symmetry 
\begin{equation}
Y_{kq}(\theta,\phi)=(-1)^{k}Y_{kq}(\pi-\theta, \pi+\phi)
\end{equation}
of the spherical harmonic functions, leads to the  
 $P,\ Q,\ {\rm and}\  F$ distribution functions in the classical limit as, 
\begin{eqnarray}
\lim_{s\rightarrow\infty}P(\theta_{1},\phi_{1};\theta_{2},\phi_{2})=&
\frac{1}{4\pi}\delta(\cos\theta_{1}+\cos\theta_{2})\ 
\delta(\phi_{1}-(\phi_{2}+\pi)),\cr
\lim_{s\rightarrow\infty}Q(\theta_{1},\phi_{1};\theta_{2},\phi_{2})=&
\frac{1}{4\pi}\delta(\cos\theta_{1}+\cos\theta_{2})\ 
\delta(\phi_{1}-(\phi_{2}+\pi)),\cr
\lim_{s\rightarrow\infty}F(\theta_{1},\phi_{1};\theta_{2},\phi_{2})=&
\frac{1}{4\pi}\delta(\cos\theta_{1}+\cos\theta_{2})\ 
\delta(\phi_{1}-(\phi_{2}+\pi)).
\end{eqnarray}
Observe that in the classical limit, these distribution functions 
  reflect  the  perfect anticorrelation between the  classical 
vectors $\vec{n}_{1}$ and $\vec{n}_{2}$.
\begin{figure}
\begin{center}
\input{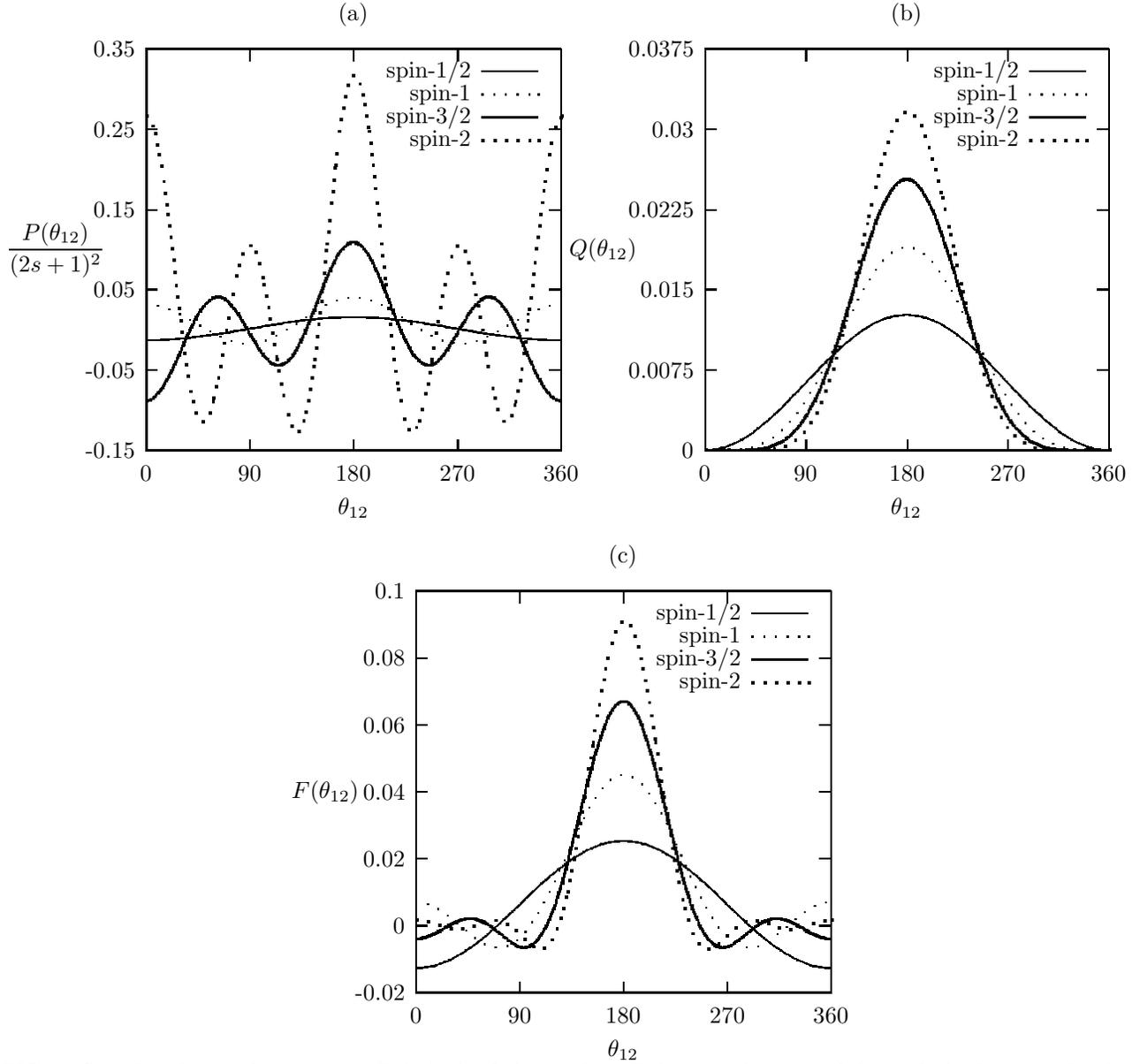}
\caption{Spin distribution functions 
(a) $P(\theta_{12})$, (b) $Q(\theta_{12})$ and (c) $F(\theta_{12})$ 
as a function of the angle $\theta_{12}$ 
between the classical spin vectors $\vec{n}_{1}$ and $\vec{n}_{2}$ 
constituting the spin singlet.}   
\end{center}
\end{figure}
\noindent{\bf ACKNOWLEDGEMENTS:} The support of IMSc(Institute of Mathematical 
Sciences, Chennai)  through the award of an Associateship is greatfully 
acknowledged. 


\begin{references}
\bibitem{multipoles}
Agarwal[7] has derived $P-,\ Q-$ functions  using the 
state multipole operators $\hat{T}^{k}_{q}$ which differ from $\t$ used here, 
by 
a factor $\sqrt{2s+1}$  i.e.,    $\hat{T}^{k}_{q}=\frac{1}{ \sqrt{2s+1}}\,\t$ 
and consequently,  
the state multipoles $\rho^{k*}_{q}$ used in Ref.[14], [15], differ 
from the Fano statistical tensors $t^{k}_{q}$  by $\rho^{k*}_{q}=
\frac{t^{k}_{q}}{\sqrt{2s+1}}.$
\bibitem{Rose3}
The factor $\displaystyle\frac{1}{k!\ {\cal N}_{sk}}$ appears because 
$\hat{\tau}^{k}_{q}
(\hat{\vec{\rm S}})$ are nothing but symmetrised solid harmonics involving 
$\hat{\vec{\rm S}}$  with a  normalisation factor[9] 
$\displaystyle\frac{1}{k!\ {\cal N}_{sk}}$.
\end{references}
\end{document}